\documentclass[conference]{IEEEtran}
\IEEEoverridecommandlockouts
\usepackage{cite}
\usepackage{amsmath,amssymb,amsfonts}
\usepackage{algorithmic}

\usepackage{wrapfig}
\usepackage{color}

\usepackage{graphicx}

\usepackage{textcomp}
\usepackage{xcolor}
\usepackage{tabularx}
\usepackage{comment}

\usepackage{tikz}
\usepackage{textcomp}
\usepackage{hyperref}
\usepackage{lipsum}

\def\BibTeX{{\rm B\kern-.05em{\sc i\kern-.025em b}\kern-.08em
    T\kern-.1667em\lower.7ex\hbox{E}\kern-.125emX}}

\def\ThreeDivisionSize{0.33}

\newcommand\copyrighttext{%
\footnotesize \textcopyright 2021 IEEE.  Personal use of this material is permitted.  Permission from IEEE must be obtained for all other uses, in any current or future media, including reprinting/republishing this material for advertising or promotional purposes, creating new collective works, for resale or redistribution to servers or lists, or reuse of any copyrighted component of this work in other works.
DOI: 10.1109/IIAI-AAI53430.2021.00011}
\newcommand\copyrightnotice{%
\begin{tikzpicture}[remember picture,overlay]
\node[anchor=south,yshift=10pt] at (current page.south) {\fbox{\parbox{\dimexpr\textwidth-\fboxsep-\fboxrule\relax}{\copyrighttext}}};
\end{tikzpicture}%
}

\begin{document}

\title{
    Developing a Lecture Video Recording System \\ Using Augmented Reality
}

\author{\IEEEauthorblockN{Yuma Ito, Masato Kikuchi, Tadachika Ozono, Toramatsu Shintani}
\IEEEauthorblockA{\textit{Department of Computer Science, Graduate School of Engineering}\\
\textit{Nagoya Institute of Technology}\\
Gokiso-cho, Showa-ku, Nagoya, Aichi, 466-8555 Japan \\
E-mail: \{higashi,kikuchi,ozono,tora\}@ozlab.org}}

\maketitle
\copyrightnotice

\begin{abstract}

Assistive technology is a prerequisite for making a high-quality lecture video.
It is therefore imperative to edit the lecture video after recording.
In this study, we aim to reduce the cumbersome task of lecture video editing by developing a system that enables the addition of visual effects in the video while recording.
In particular, we use augmented reality (AR) technology to digitize and display in real-time lecture materials, assistant agents, and other recording contents used by the lecturer.
Our system realizes such a mechanism as a lecture recording environment.
In addition, our system based on AR technology can support the work of the lecturer, which is difficult to do by oneself while conducting the lecture, using the information of the lecturer's position and the progress of the lecture.
We evaluated the system functionality and performance, and verified the system's correct behavior. 
If the burden of making lecture videos can be reduced, the lecturer will be able to devote more time to improving the quality of lecture contents, which is expected to contribute to the improvement of lectures.

\end{abstract}


\section{Introduction}
\label{sec:Introduction}

\renewcommand{\thefootnote}{\roman{footnote}}
In some cases, lecture videos are used as pre-distributed materials in on-demand classes and flipped classrooms\cite{response_collector}.
Lecturers produce lecture videos in various ways.
The lecture video that this study aims to produce shows the lecturer's facial expressions and motions, is easy to watch, and has visual effects to a certain extent.
Lecture videos are preferred to be in a format that shows the lecturer\cite{video_effect}.

Making lecture videos consists of recording and editing. 
After the recording, lecturers usually edit videos to add rich effects, however, editing is time-consuming. 
We aim to realize a system that can automatically add appropriate effects to a video while recording it. 
This means that lecturers need not edit after recording to create high-quality lecture videos.
To reduce editing work after recording, this system must operate exactly as intended by the user while recording a long lecture video.

Our approach provides support for video making during lecture recording, thereby reducing the burden after recording.
Specifically, our system uses augmented reality (AR) technology to create a lecture video with added visual effects in real-time in accordance with the progress of the lecture.
Our system digitizes and displays lecture materials, lecture assistant agents, and other lecture recording contents.
The lecturer uses these tools to conduct lectures in augmented reality (AR) space.
Our system provides the lecturer such a lecture video recording environment in the AR space. 
Our system using digital lecture contents can adapt to various lecture formats envisaged by lecturers, even when recording a lecture without the need for video editing.
Another advantage of using AR technology, which is a feature of our system, enables our system to handle three-dimensional information based on depth data.
For instance, it includes the plane information and position of the lecturer in the lecture space.
Our system uses this information to configure the display of AR Slide, to control lecture assistant agent, and system functions for lecture support. 
Especially in designing such a system, we need to reduce unwanted editing work after recording due to system errors or delays.

The rest of the paper is organized as follows. Section \ref{sec:Related Work} presents related works. 
Section \ref{sec:Lecture Video Making Support} describes how to support the creation of lecture videos using AR technology. 
Sections \ref{sec:Lecture Video Recording System} and \ref{sec:Evaluation} describe the implementation and experiments. 
Finally, we discuss our system considering the results in Section \ref{sec:Discussion} and conclude this paper in Section \ref{sec:Conclusion}.

\section{Related Work}
\label{sec:Related Work}
\subsection{Video Effect Production Support}

\begin{figure*}[tb]
    \begin{tabular}{ccc}
      \begin{minipage}[t]{\ThreeDivisionSize\textwidth}
        \begin{center}
          \includegraphics[keepaspectratio, width=\hsize]{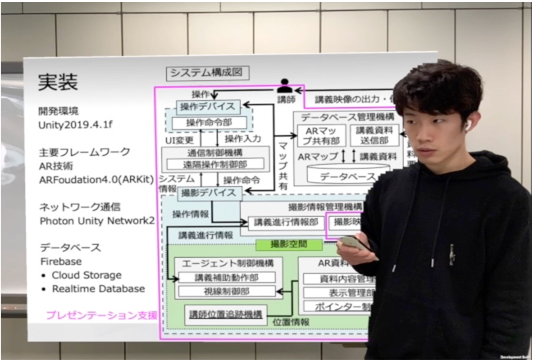}
          \hspace{1.6cm} (a) one AR Slide
        \end{center}
      \end{minipage}
      \begin{minipage}[t]{\ThreeDivisionSize\textwidth}
        \begin{center}
          \includegraphics[keepaspectratio, width=\hsize]{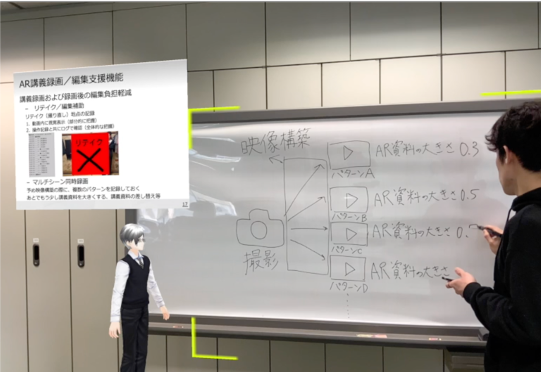}
          \hspace{1.6cm} (b) real lecture material (whiteboard)
        \end{center}
      \end{minipage}
      \begin{minipage}[t]{\ThreeDivisionSize\textwidth}
        \begin{center}
          \includegraphics[keepaspectratio, width=\hsize]{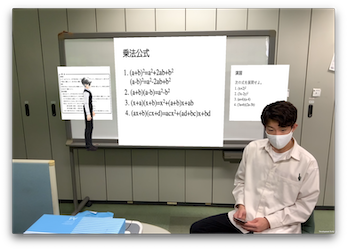}
          \hspace{1.6cm} (c) multi AR Slides
        \end{center}
      \end{minipage}
    \end{tabular}
    \caption{Example of displayed lecture material by our system.}
    \label{fig:slidemode_example}
  \end{figure*}

Some researchers have focused on adding visual effects to lecture and presentation videos using AR and 2DCG technology to have a high effect in terms of expression and entertainment. In particular, the addition of visual effects using AR technology is superior in that it enables three-dimensional visual expression and reduces complicated video editing work.

In a previous study, we developed a video editing system for flipped classrooms using AR technologies\cite{lecture_videomake_ARpuppet}.
We designed the agents playing the role of a virtual student in AR space, called AR puppets, which interact with the lecturer and help students learn.
We used AR puppets and created lecture videos together to add visual effects to the lecture videos.
Saquib et al.\cite{Interactive_body-drive} developed a system that displays materials such as figures and graphs in videos during a live presentation, with which the presenter can interact in real time with gestures and postures.
Relevant to our system is to display the digital materials and the presenter as if they were in real space.
It allows the audience to watch it as a single video content without having to pay attention to both materials and the presenter. 

Many lecturers create lecture videos using standard presentation tools such as Microsoft PowerPoint\footnote{https://www.microsoft.com/en-us/microsoft-365/powerpoint}and Keynote\footnote{https://www.apple.com/jp/keynote/}.
These tools make it easy to create a narrated presentation or presentation with an overlaid lecturer's capture.
Our study focuses on features of AR technology, which realize added graphical expressions of presentation using three-dimensional spatial information to video during lectures without any editing work.
Our system uses digital lecture contents for lectures, such as lecture materials and agents.
This can flexibly adapt to various lecture formats by manipulating and adjusting the lecture recording contents (Fig. \ref{fig:slidemode_example}).

\subsection{Human-like Virtual Agent}

Human-like virtual agents, which have a form and communication similar to humans, have been researched and used in various situations, as language instructors\cite{language_trainers_agent} and a museum guide\cite{museum_agent}.
Particularly in a presentation situation, the presenter makes a collaborative presentation with a virtual agent\cite{co_presentation}, and projects an avatar that reflects the presenter's behavior on the slides\cite{embodied_interactions}.
Mumm et al.\cite{AR_agent_effect} investigated how the presence of virtual agents can affect the motivation of participants toward given tasks.
It is inferred that abstracted humanoid agents may have a positive effect on human motivation.
We notably focus on visual effects of virtual agents without the need for manpower.

In this study, we use a human-like virtual agent as the lecture assistant agent, who provides the graphical effect as video content and lecture assistance for making the lecture videos more comprehensible.
In a previous study, we explored ways to control a virtual agent in an AR space\cite{ARPopupPictureBook}.
We considered whether the performer could control the virtual agent himself or herself without disturbing the progress of performance.
We extend this work further and design three-dimensional control of lecture assistant agent.
Our system controls lecture assistant agent semi-autonomously using the information in the lecture space, such as the lecturer's position, lecture materials, and lecture progress.

\section{Lecture Video Making Support using AR technology}
\label{sec:Lecture Video Making Support}

Our system realizes a collaborative presentation by a lecturer and lecture assistant agent in the AR space.
We assume that the lecturers record lecture videos by themselves, and the length of typical videos is longer than 30 minutes.
This section describes in detail how we support the creation of lecture videos using AR technology.

\subsection{Design Lecture Material Object}

Lecturer can use two types of lecture materials in our system: AR Slide and real lecture materials.
AR Slide is a lecture material displayed as a digital object in the AR space using AR technology.
AR Slide can be used in image formats such as PowerPoint, textbooks, and reference images, as well as in video format, taking into account the lecture format assumed by each lecturer.
The real lecture materials in this paper refer to the materials that the lecturer writes directly on real objects such as blackboards and whiteboards.
The lecturer usually gives a lecture using this AR Slide.

By using AR Slide, there is no need for a monitor or projector to display the lecture materials, nor is there a need to edit the video later, and it is possible to display the materials following the progress of the lecture.
In addition, it can be displayed clearly.
A digitally displayed AR slide can be displayed stably without being affected by the actual recording environment.
Moreover, AR technology has people occlusion, which grasps the front-back relationship between AR objects and people and renders AR contents appropriately.
This technology makes it easy to realize representations that look like chroma key compositing of AR Slide and the lecturer.

The type of lecture videos that different lecturers want to make varies.
This may also vary depending on the field and content of the lecture.
It would be a burden for the lecturer to prepare for the lecture video that the lecturer assumes in advance or in the middle of the recording.
Our system enables the lecturer to freely manipulate and adjust the AR Slide, and corresponds to various lecture formats that change the display style of AR Slide.
The layout of the materials is limited with a monitor or projector screen.

AR technology makes it possible to place the lecture materials anywhere in AR space.
The lecturer can freely adjust the size and aspect ratio of AR Slide.

There are mainly three display styles of AR Slide.
Fig. \ref{fig:slidemode_example} denotes the actual display styles of AR Slide by our system.
(a) is the most basic display style, showing a single slide.
(b) is the case when the lecturer uses real lecture material such as a blackboard and whiteboard.
The AR slide is moved to the upper left corner, and the lecturer gives a lecture using a real lecture material.
(c) is a style in which multiple slides are displayed in the video.
Fig. \ref{fig:slidemode_example}c shows the main slide and previous and next slides are displayed.

In another case, the lecturer can proceed with the lecture displaying the left slide fixedly on a particular slide (this function is called a Pin Function).
The lecturer can switch these styles appropriately during recording, allowing the lecturer to flexibly change depending on the content of the lecture.

\subsection{Lecture Assistant Agent}

Our study realizes an agent provides supplementary explanations for lectures.
Lecture assistant agent plays two roles.
One is to provide a visual effect to relieve the monotony of the lecture video.
The other is the role of assisting the lecture.
We are familiar with the format in news and commercials, where the main speaker and assistant speaker are separated and give a presentation together.

Because the lecturer records a lecture video by himself or herself, the lecture assistant agent should be controlled semi-autonomously.
Proceeding with the lecture itself becomes a burden to the lecturer.

Therefore, it is necessary to avoid that the control of the lecture assistant agent becomes obstructive to the progress of the lecture.
In our system, lecture assistant agent semi-autonomously controls its movements using comprehensive information about lecture progress, lecture materials, the position of the lecturer and pointer, and so on.

\begin{figure}[!tb]
  \begin{center}
    \includegraphics[keepaspectratio, width=\hsize]{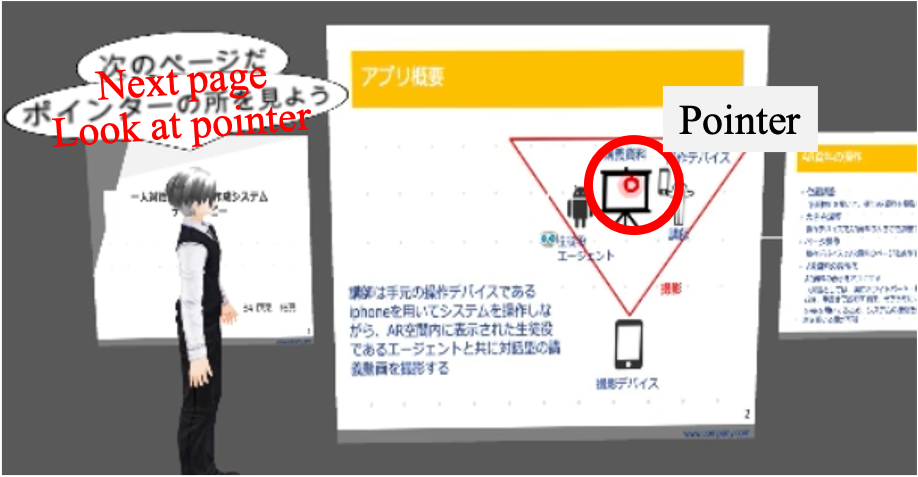} 
    \caption{Visual Effects by lecture assistant agent. The agent pop-up AR objects in a text format chronologically above its head.}
    \label{fig:example_agent}
  \end{center}
\end{figure}

We assumed that there would be two types of supplementary explanations needed in a lecture:
explanations about the lecture content and about the lecture progress. 
Supplementary explanations about the lecture content require appropriate content and timing for the progress of the lecture.
This is because meaningless explanations can distract students from understanding lecture video content.
Because it is difficult for the agent to provide such explanations with semi-autonomous control, this paper mainly focuses on supplementary explanations about the lecture progress.
Fig. \ref{fig:example_agent} denotes lecture assistant agent that prompts the viewer to pay attention when the lecture progress transitions, such as switching slides or using a pointer.

\subsection{Functions using Lecturer's Recording Information}

\begin{figure}[tb]
  \begin{center}
    \includegraphics[keepaspectratio, width=\hsize]{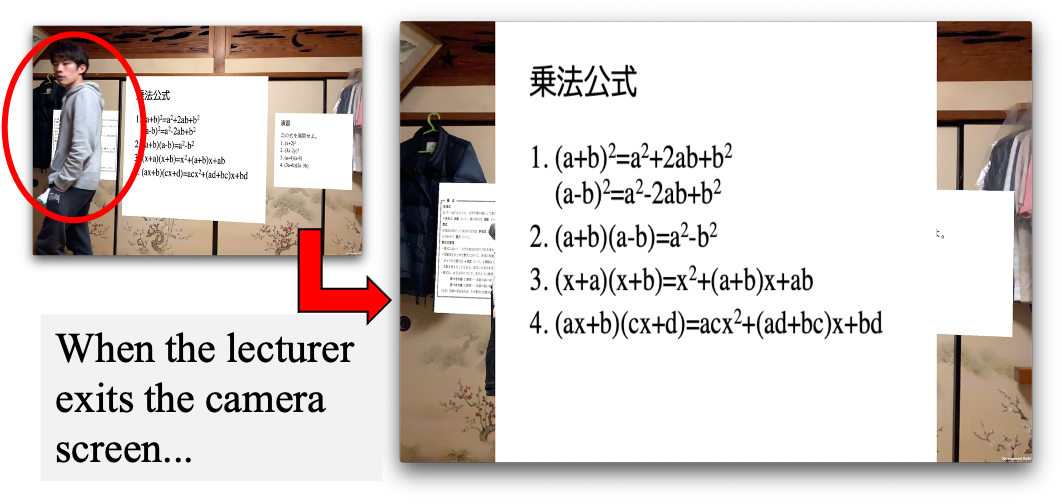} 
  \end{center}
  \begin{center}
    \includegraphics[keepaspectratio, width=\hsize]{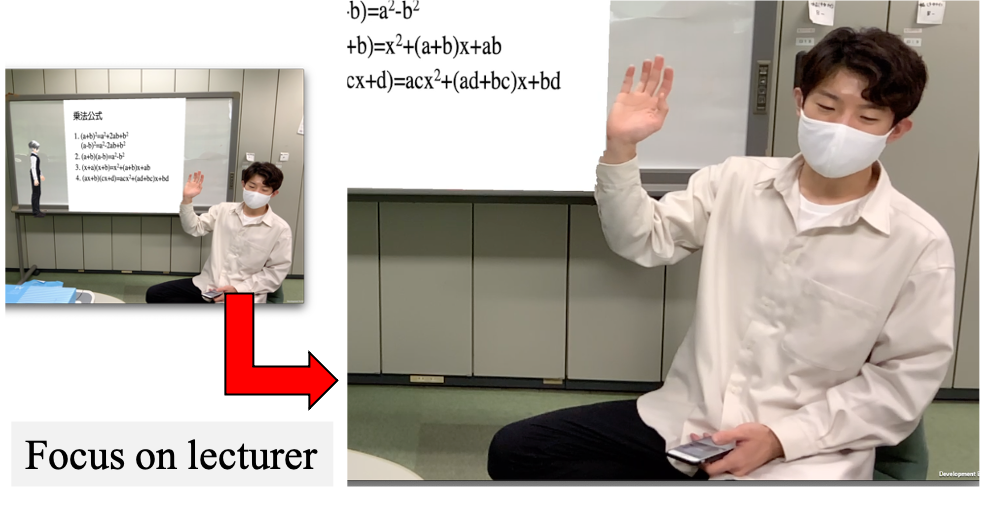} 
  \end{center}
   \caption{
     Recording Content Switching Function.
     It is automatically controlled camera using the lecturer's position information.
     Above: Close-up display of lecture materials. Below: Close-up display of the lecturer.
   }
   \label{fig:example_recordingContentsToggle}
\end{figure}

\begin{figure}[tb]
  \begin{center}
    \includegraphics[keepaspectratio, width=\hsize]{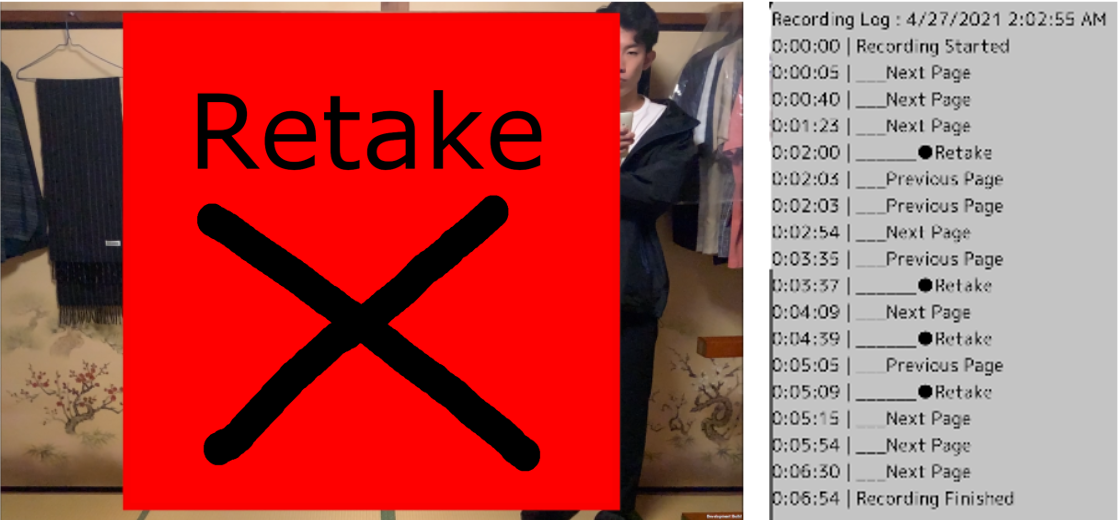} 
    \caption{
        Video cut editing assistance function after lecture recording.
        Two methods to save the retake point are used.
        Left: Insert a visual effect into the video.
        Right: Save as a text log along with the operation record.
    }
    \label{fig:retake}
  \end{center}
\end{figure}

Our system handles two types of information as Lecturer's Recording Information: the lecturer's position and the progress of the lecture.
On TV news, we can see that the camera crews switch between the material, the newscaster, and the guest according to the situation. 
It is possible that adjusting recording contents according to the progress of the lecture will make it easier for viewers to see and understand the contents.
However, it is desirable to avoid direct control of the camera, which requires human work, and to make it as automatic as possible.
An example of the recording contents switching function of this system is shown in Fig. \ref{fig:example_recordingContentsToggle}.
This function provides a close-up view of the lecture materials when the lecturer exits the camera screen (The above Fig. \ref{fig:example_recordingContentsToggle}).
In addition, our system controls the camera based on the lecturer's position and displays a close-up of the lecturer (Fig. \ref{fig:example_recordingContentsToggle} below).

Our system creates lecture videos with added visual effects in real time to reduce the burden of video editing.
However, editing video after recording is necessary to some extent, such as cut editing for retake parts.
Therefore, we propose a system to record the retake point of the video using the information of the lecture's progress and to assist in editing the video after recording.
Fig. \ref{fig:retake} denotes two recording methods: one is to record by inserting a visual effect into the video, and the other is to record by saving as text log along with the operation log.

\section{Lecture Video Recording System}
\label{sec:Lecture Video Recording System}

\begin{figure}[tb]
  \begin{center}
    \includegraphics[keepaspectratio, width=\hsize]{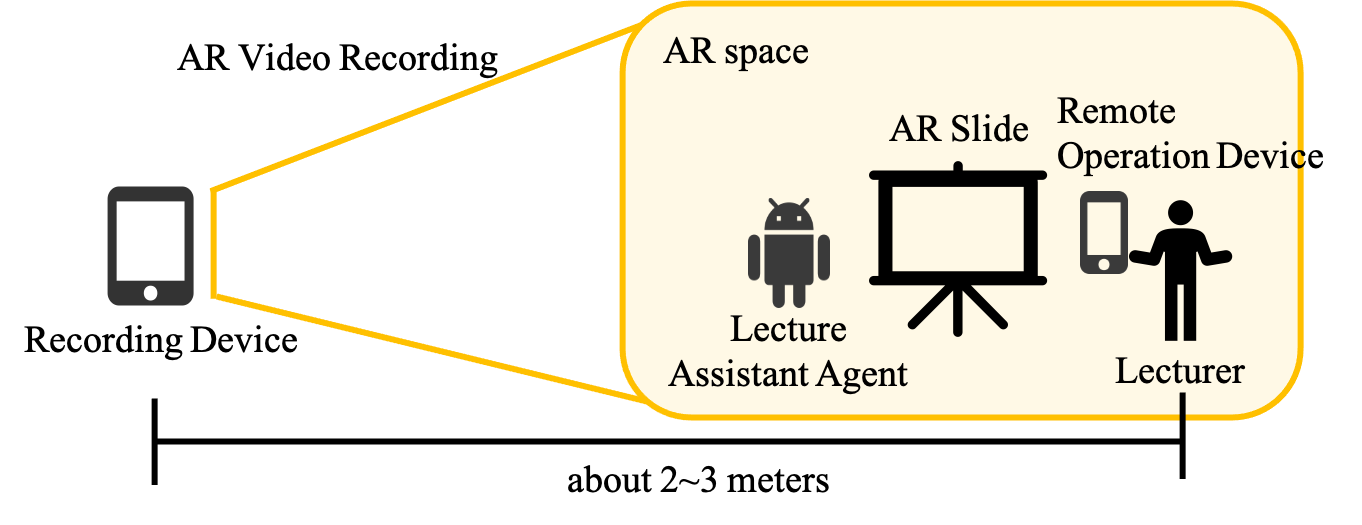} 
    \caption{Hardware setup.}
    \label{fig:hardware}
  \end{center}
\end{figure}

Our system, developed as iOS application with Unity, uses two iOS devices.
Fig. \ref{fig:hardware} denotes the hardware configuration of our system.
These devices are called recording device and remote operation device, respectively.
Recording Device creates lecture videos with added visual effects.
Because the lecturer is approximately several meters away from the recording device, it is impossible to directly control the recording device during the lecture.
Therefore, the recording device, that is, the recording system, was remotely controlled via a remote operation device.
We used iPhone 8 running on iOS14.4 as the remote operation device and iPad Pro 2nd generation running on iOS14.4 as the recording device.

\subsection{System Flow}

We describe the procedure for making a lecture video with our system. 
The procedure consists of four steps mentioned as follows: 1) launch the application, 2) register the recording device and remote operation device, 3) prepare for recording: track lecturer's position and put AR slide with recording device, and 4) record a lecture: conduct a lecture with a remote operation device.

In first step, the lecturer starts this application on each of the two iOS devices.
After some time, the network is automatically set up between the two devices to create a communication environment.
In second step, each device is assigned to a recording device or a remote operation device.

The lecturer prepares for the lecture recording in third step.
Specifically, the lecturer sets up lecturer's position tracking and AR Slide.
The lecturer always holds a remote operation device while recording because of the specifications of this system.
Therefore, our system estimates the lecturer's position based on the position coordinates of the remote operation device.
Therefore, it is necessary to acquire the position coordinates of the remote operation device in the AR space created by the recording device.
This mechanism is implemented by a remote operation device sharing its coordinates with a recording device through self-localization estimation in AR technology.
However, the coordinate information provided by the self-localization estimation is expressed in relative coordinates from the origin of the AR space constructed by oneself.
Hence, it is essential to match the origin and axis (direction) of the AR space that each device constructs.

Our system uses the overall spatial mapping information, which is referred to as the AR map, and shares it among other devices to align the origin and axis of the AR space.
Subsequently, it estimates the position of the remote operation device and lecturer by self-localization estimation.

\begin{figure}[tb]
  \begin{center}
    \includegraphics[keepaspectratio, width=\hsize]{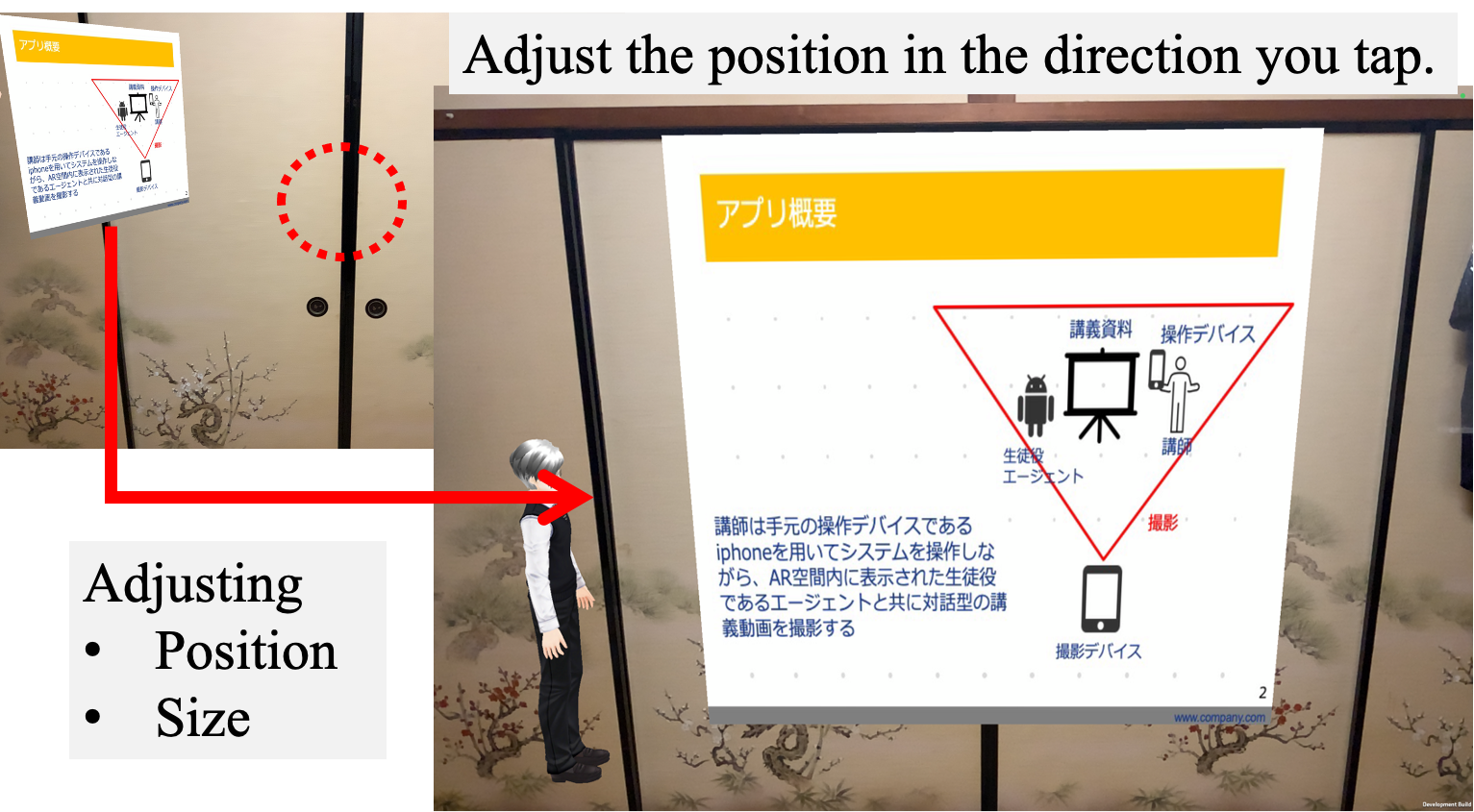} 
    \caption{Adjusting function of AR Slide.}
    \label{fig:adjustARDoc}
  \end{center}
\end{figure}

When setting up the AR slide, the system displays the lecture materials prepared in advance as AR Slide.
In addition, the lecturer can adjust the position and size of the AR slide (Fig. \ref{fig:adjustARDoc}).
Our system uses the plane detection of AR technology to render the lecture materials as if they were pasted on a vertical surface in the recording space.

After finishing preparation for recording, the lecturer can give a lecture by manipulating the AR slide.
The lecturer will use the system functions according to the content and progress of the lecture.
In the previous section, we introduced some system functions.

\begin{figure}[tb]
  \begin{center}
    \includegraphics[keepaspectratio, width=\hsize]{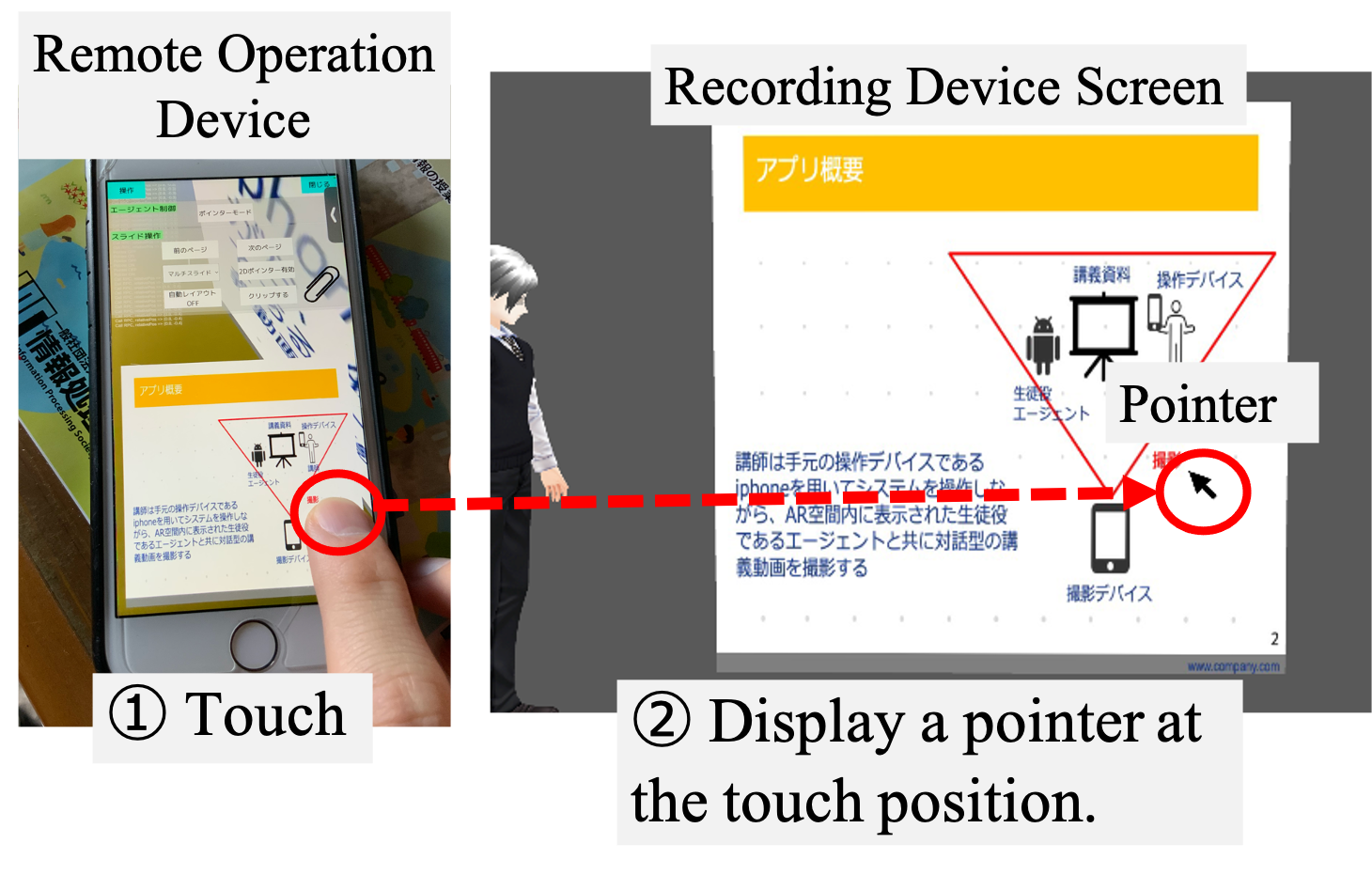} 
    \caption{
      Pointer function.
      The remote operation device displays a 2D image of a material synchronized with an AR slide that exists in the AR space of the recording device.
    A pointer is displayed in relation to the touched position of the material displayed on the remote operation device.}
    \label{fig:pointer}
  \end{center}
\end{figure}

The other function of our system is to display a pointer on the AR slide (Fig. \ref{fig:pointer}).

\subsection{System Architecture}

\begin{figure}[tb]
  \includegraphics[keepaspectratio, width=\hsize]{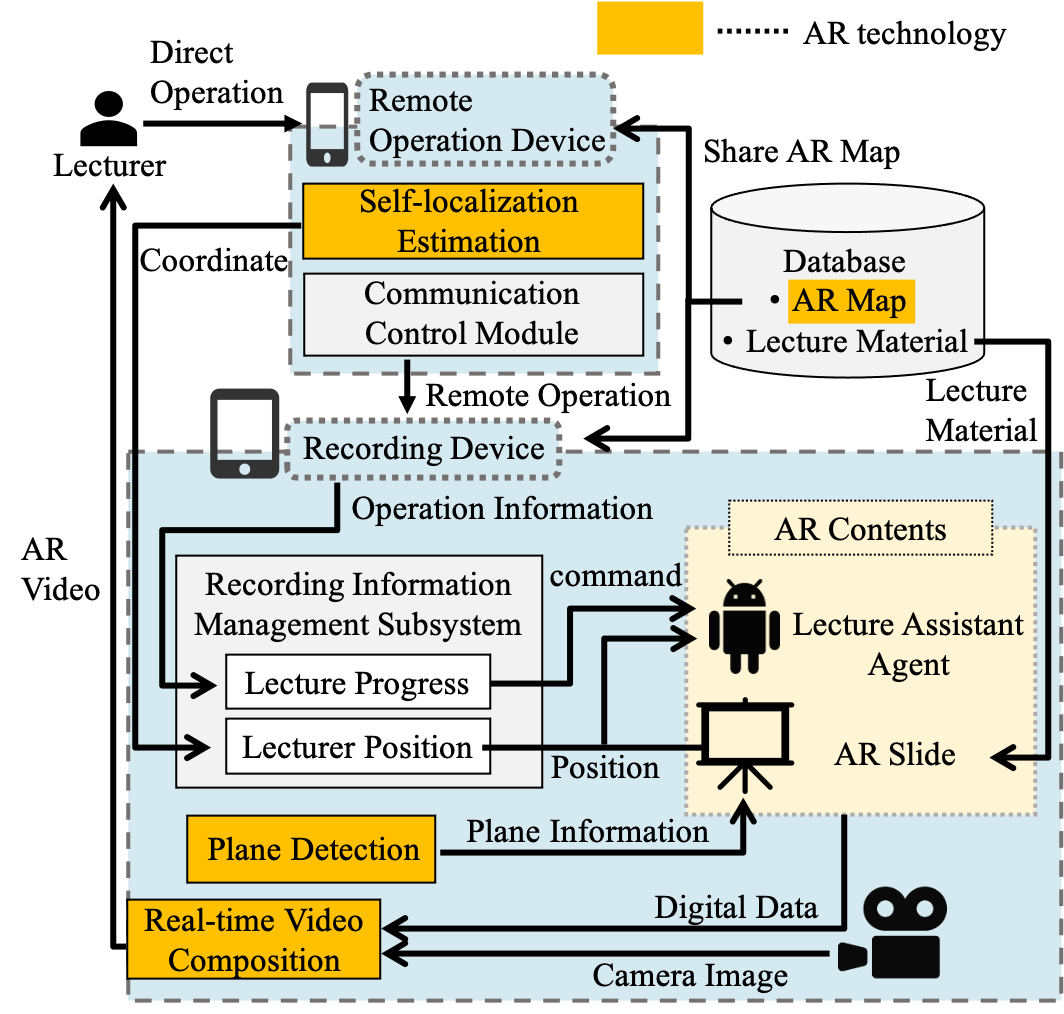} 
  \caption{System Architecture.}
  \label{fig:system_diagram}
\end{figure}

Our system consists of a recording device, a remote operation device, and a database (Fig. \ref{fig:system_diagram}).

The remote operation device has a communication control module that is responsible for the communication function between devices.

The lecturer remotely controls the recording device via the communication control module based on the input from the remote operation device. 
This module also updates the UI of the remote operation device by receiving the system information from the recording device, such as lecture information and results of operation commands, so that the lecturer can directly receive the system feedback.

The database holds two types of data: AR map and lecture material.

The system needs to share an AR map with each device to track the lecturer's position. 
Specifically, one device sends the AR map to the other device via the database.
The lecture materials for AR slide are also stored in the database.
The material is saved in the database as image or video format.
The lecturer can freely change and use the materials from a web browser by referring to them from the database without the need to import them into the device in advance.

Recording device mainly comprises recording information management subsystem, lecture assistant agent, and AR slide.
The recording information management subsystem handles the lecture progress and lecturer's position information as recording information.
Since the lecturer uses AR Slide and system functions to conduct a lecture, the progress of the lecture is directly related to the system's operation.
Therefore, the system determines the lecture progress based on the operation information.

Lecture assistant agent is controlled semi-autonomously.
Recording information management subsystem commands the agent to perform lecture assistance considering lecture progress.
The system controls the gaze and posture of the lecture assistant agent in three dimensions, depending on the position of the lecturer and AR Slide.
AR Slide is displayed by projecting material data onto a flat AR image object.
AR Slide has a layered structure of multiple AR objects.
The system controls the AR Slide, such as page manipulation and changing display style, by a combination of showing and hiding these objects and adjusting their position and size.
Finally, the system makes AR video as a lecture video in real time by combining digital data of AR contents and camera images using AR technology.

\section{Evaluation}
\label{sec:Evaluation}

In this section, we evaluate the functionality and performance of our system, as a recording system using AR technology.
We conducted experiments to demonstrate that our system can add appropriate visual effects in real time in accordance with the lecturer's operation. 
Specifically, we measured their accuracy and process speed.
We evaluated that the system can operate exactly as the user intended while recording a lecture video.
The system should perform appropriate editing in real time for the lecturer's operations.
In other words, the system can select appropriate commands for the lecturer's operations.
Therefore, we measured the elapsed time as the response speed of the system, which is from the issue of command by the remote operation device to the completion of that system function.
We selected four functions, which are required to execute in real time.
We measured 100 times for each function and calculated the average and maximum values.
Table I shows that the accuracy is 100\% and response time is low.
The average and maximum response time for the Pin Function is 74 ms and 302 ms, respectively, which is the slowest among the measured functions.
The results indicate that our system has sufficient real-time performance for adding visual effects during lecture recording.

Our system allows the use of the information in the recording space, such as the lecturer's standing position.
As explained in Section \ref{sec:Lecture Video Recording System}, the accuracy of the lecturer's position recognized by the system depends on the position tracking accuracy of the remote operation device.
Therefore, we measured the accuracy of the position tracking of remote operation device.
We describe the measurement procedure.
First, the system places a target object randomly in the AR space of the recording device.
Second, the user overlaps remote operation device on the target object.
Finally, the system measures the distance between the position of the target object and the tracked position of the remote operation device by the system.
Because there might be some hand tremors and misrecognition, the coordinates by position tracking are calculated as a five-second moving average value.
We conducted the experiments in two cases.
The results are shown in Table \ref{tb:result_sharelocation}.
Case 1 involves measuring in various spaces within a short period after setting up tracking.
Case 2 involved setting up tracking only once and measuring it in the same space for 30 min.

\begin{table}[htb]
  \centering
  \caption{Response time and accuracy}
  \begin{tabular}{|l||c|c|c|} \hline
    & max & mean & accuracy\\ 
    & [ms] & [ms] & [\%] \\ \hline \hline
    Page Operation & 189 & 61 & 100 \\ \hline
    Display Format Change & 162 & 53 & 100 \\ \hline
    Pointer & 177 & 69 & 100 \\ \hline
    Pin Function & 302 & 74 & 100 \\ \hline
  \end{tabular}
  \label{tb:result_systemSpeed}
\end{table}

\begin{table}[htb]
  \centering
  \caption{Position error of recognized Remote Operation Device}
  \scalebox{0.9}{
  \begin{tabular}{|l||c|c|c|c|c|} \hline
    & count & min & max & mean & variance \\ 
    &  & [cm] & [cm] & [cm] & \\ \hline \hline
    \begin{tabular}{l}
    Case 1
    \end{tabular}
    & 100 & 2.5 & 35.5 & 16 & 60.7 \\ \hline
    \begin{tabular}{l}
    Case 2
    \end{tabular}
    & 54 & 3 & 35.6 & 12.7 & 44.7 \\ \hline
  \end{tabular}
  }
  \label{tb:result_sharelocation}
\end{table}

\section{Discussion}
\label{sec:Discussion}

Our system provides an environment for recording lecture videos in AR space.
As shown in Table \ref{tb:result_systemSpeed}, to reduce the burden of video editing, the system adds visual effects to the video in real time.
Our experimental results show that the system can add appropriate visual effects for the lecturer's operations; therefore, the system does not cause unwanted editing work after recording due to system errors or delays. 
Our system can help lecturers create lecture videos because the system can add visual effects to lecture videos on recording without time-consuming editing work. 
Furthermore, the system is also expected to reduce the additional work, such as retaking and video editing after recording due to the addition of visual effects beyond the lecturer's expectations.

The feature of our system is the acquisition of spatial information using AR technology, and the utilization of the same to create visual effects and functions.
The accuracy of the acquired spatial information affects the functionality of the entire system.
In Case 1, the average error of the position tracking of the remote operation device was 16cm. 
From a system perspective, tracking accuracy is sufficient, even to a certain degree, rather than a strict accuracy.
Therefore, the performance of the system is practical enough.

We assume that a lecturer records long lecture videos by using our system, so the system is required to execute appropriate commands for the lecturer's operations in real-time in order to reduce time-consuming editing work after long recording.
Because of the specification of AR technology, the extraction of spatial feature points is always executed as long as the system is running.
Therefore, it is assumed that the accuracy will decrease as the system is working for a long time because of the specification of position tracking.
However, we did not observe any loss in accuracy (Case 2).
The variance was smaller than that observed in Case 1.
The results indicated that the newly acquired spatial feature points were processed appropriately without conflicting with the old.
We conclude that the tracking implementation of our system is practical for long-term lecture recording.

\section{Conclusion}
\label{sec:Conclusion}

We proposed a lecture video recording system providing AR space for giving lectures including AR Slide, Lecture Assistant Agent, and so on. 
Our experimental results demonstrated that the system can add visual effects to lecture videos on recording without unwanted editing work caused by system errors or delays. 
Our system can help lecturers create lecture videos with visual effects to help students.
However, we need to conduct usability tests of our system in the near future.

\section*{Acknowledgment}
This work was supported in part by JSPS KAKENHI Grant Numbers 19K12097,19K12266.

\vspace{12pt}
\color{red}

\end{document}